# Analytical model for predicting folding stable state of bistable deployable composite boom

Tian-Wei Liu[1,2,3], Jiang-Bo Bai[1,2]*, Nicholas Fantuzzi[3]

1. School of Transportation Science and Engineering, Beihang University, Beijing, 100191, People's Republic of China (*, corresponding author: baijiangbo@buaa.edu.cn)

2. Jingdezhen Research Institute of Beihang University, Jiangxi Province, 333000, People's Republic of China

3. DICAM Department, University of Bologna, Bologna 40136, Italy

**Abstract:** The bistable deployable composite boom (Bi-DCB) can achieve bistable function by storing and releasing strain energy, which has a good application prospect in space field. For example, it serves as the main support section of deployable structures (e.g., solar arrays and antennas). This paper investigates the folding stable state of the Bi-DCB through the analytical method. Based on the classical Archimedes' helix, the geometrical model of the Bi-DCB was established. Using energy principle, an analytical model for predicting the folding stable state of the Bi-DCB was presented. The failure indices of six Bi-DCBs in the folding stable state were calculated using the Tsai-Hill criterion and the maximum stress criterion. To validate the analytical model proposed in this paper, the prediction results were compared with the results of two Finite Element Models (FEMs) and experimental results, and the four were in good agreement. Finally, the effect of geometric parameters (i.e., radius of cross-section, thickness and length) on the folding stable state of the Bi-DCB was further investigated with the aid of the analytical model. It is shown that geometric parameters are one of the key factors affecting the folding stable state of the Bi-DCB.

**Nomenclature**

| | |
|---|---|
| $a$ | controls the distance between two adjacent circles |
| $b$ | distance from the start-point to the origin of the polar coordinate system |
| $E_1$ | longitudinal elastic modulus of the ply, GPa |
| $E_2$ | transverse elastic modulus of the ply, GPa |
| $G_{12}$ | in-plane shear modulus, GPa |
| $H_1$ | transformation variable |
| $H_2$ | transformation variable |
| $I_{f,1}$ | Tsai-Hill failure index |
| $I_{f,2}$ | maximum stress failure index |
| $L$ | length of the Bi-DCB, mm |
| $M_x$ | internal bending moment per unit length on the cross-section of the laminate, N |
| $M_y$ | internal bending moment per unit length on the cross-section of the laminate, N |
| $M_{xy}$ | internal bending moment per unit length on the cross-section of the laminate, N |
| $N_x$ | internal force per unit length on the cross-section of the laminate, N/mm |
| $N_y$ | internal force per unit length on the cross-section of the laminate, N/mm |
| $N_{xy}$ | internal force per unit length on the cross-section of the laminate, N/mm |
| $r_0$ | polar radius at the start-point of the Bi-DCB in the folding stable state, mm |
| $r_1$ | polar radius at the end-point of the Bi-DCB in the folding stable state, mm |
| $R$ | radius of cross-section, mm |
| $S_{12}$ | in-plane shear strength of composite ply, MPa |
| $t$ | thickness of the Bi-DCB, mm |
| $u$ | strain energy per unit area, mJ /mm |
| $U$ | total strain energy for a unit longitudinal length of the Bi-DCB in the folding stable state, mJ |
| $X_t$ | longitudinal tensile strength of composite ply, MPa |
| $X_c$ | longitudinal compressive strength of composite ply, MPa |
| $X_1$ | longitudinal strength of composite ply, MPa |
| $X_2$ | longitudinal strength of composite ply, MPa |
| $Y_t$ | transverse tensile strength of composite ply, MPa |
| $Y_c$ | transverse compressive strength of composite ply, MPa |
| $Y$ | transverse strength of composite ply, MPa |
| $\alpha$ | polar angle of fitted Archimedes' helix, rad |
| $\alpha_0$ | polar angle at the start-point of Archimedes' helix, rad |
| $\alpha_1$ | polar angle at the end-point of Archimedes' helix, rad |
| $\beta$ | ply angle, deg |
| $\theta$ | central angle of the cross-section, deg |
| $\nu_{12}$ | poisson's ratio of the ply |
| $\nu_{21}$ | Poisson's ratio of the ply |
| $\sigma_1^k$ | maximum longitudinal principal stress of the $k^{th}$ ply in the laminate, MPa |
| $\sigma_2^k$ | maximum transverse principal stress of the $k^{th}$ ply in the laminate, MPa |
| $\sigma_x^k$ | maximum stress of the $k^{th}$ ply in the $x$ direction, MPa |
| $\sigma_y^k$ | maximum stress of the $k^{th}$ ply in the $y$ direction, MPa |
| $\tau_{12}^k$ | maximum principal shear stress of the $k^{th}$ ply in the laminate, MPa |
| $\tau_{xy}^k$ | maximum shear stress of the $k^{th}$ ply in the $x$-$y$ direction, MPa |
| $\Pi$ | total strain energy of the Bi-DCB in the folding stable state, mJ |
| $\rho$ | polar radius of the Bi-DCB in the folding stable state, mm |

| $\varepsilon_x$ | normal strain in the *x* direction of the laminate |
| $\Delta\varepsilon_x$ | change of normal strain in the *x* direction of the laminate |
| $\varepsilon_y$ | normal strain in the *x* direction of the laminate |
| $\Delta\varepsilon_y$ | change of normal strain in the *x* direction of the laminate |
| $\gamma_{xy}$ | shear strain in the *x-y* direction of the laminate |
| $\Delta\gamma_{xy}$ | change of shear strain in the *x-y* direction of the laminate |
| $\kappa_x$ | curvature in the *x* direction of the laminate |
| $\Delta\kappa_x$ | change of curvature in the *x* direction of the laminate |
| $\kappa_y$ | curvature in the *y* direction of the laminate |
| $\Delta\kappa_y$ | change of curvature in the *y* direction of the laminate |
| $\kappa_{xy}$ | curvature in the *x-y* direction of the laminate |
| $\Delta\kappa_{xy}$ | change of curvature in the *x-y* direction of the laminate |
| Bi-DCB | bistable deployable composite boom |
| FEM | finite element model |

## 1. Introduction

With continuously evolving engineering technology, the requirement for next-generation intelligent structures and materials with novel functions has increased significantly. As a new type of intelligent composite structure, the bistable composite structure has attracted more and more attention due to its advantages of light weight, excellent mechanical properties and high space utilization [1,2]. Bistable composite structures have two stable configurations that remain in their respective equilibrium positions without the continuous application of an external force [3]. Due to the above characteristic, bistable composite structures are applied in many fields, including aerospace

(e.g., deployable booms and morphing skins) [4-9], civil engineering [10], bionic structures (e.g., flytraps) [11], energy harvesting [12,13], robots [14], etc.

As a bistable composite structure with promising application prospects, the bistable deployable composite boom (Bi-DCB) with antisymmetric layup is utilized as the deployment mechanism of solar array, with advantages of high storage ratio and light weight [15], as shown in Fig. 1. In 1996, after the British scholar Daton Lovett first proposed the design scheme of the bistable deployable composite boom (Bi-DCB) with antisymmetric layup [16], the University of Cambridge took the lead in carrying out relevant research on this structure. Based on the classical laminate theory and the minimum energy principle, Iqbal et al. [17] established a simple linear elastic bistable analytical model. The expressions of strain energy with respect to the longitudinal and transverse curvature and the central angle of the cross-section was obtained, and the radius of the cross-section of the folding stable state was predicted. However, this analytical model can not distinguish the behaviour of symmetric and antisymmetric laminates. Galletly and Guest [18] extended the analytical model proposed by Iqbal and presented a beam model to include all possible deformation modes, including twist and shear, and stretching-bending coupling effects. The predicted radius of the cross-section of the folding stable state is in good agreement with numerical simulation results, but there is a big difference with experimental results. Later, Galletly and Guest [19] proposed a shell model, which deletes the assumption of cross-section shape. Prediction results are basically consistent with the beam model, but differ greatly from experimental results. Galletly proposes that the result disparities are due to the fact that the polypropylene matrix had entered the plastic region, while the analytical models are based on linear-elastic material assumptions. Guest and Pellegrino [20] developed a two-parameter analytical model. This model assumes that the initial configuration is stress-free and subject to plane stress under other conditions. Moreover, the mid-surface of the shell undergoes bending without stretching, ensuring that all deformations are uniform, inextensional, and the Gaussian curvature is preserved and zero. This fundamental assumption allows for the fitting of all possible shell configurations onto the surface of a cylinder. Consequently, these configurations could be defined by only two parameters: the curvature of the underlying cylinder, and the orientation angle relative to the cylinder's axial axis. Prediction results of this model are in good agreement with those of the above three analytical models [17-19].

From the previous review, it is evident that the folding stable state of the Bi-DCB (especially its geometric configuration) has been extensively investigated. However, the aforementioned analytical models [17-20] fail to accurately predict the true geometric configuration of the folding stable state of the Bi-DCB with longer lengths. In these cases, the cross-sectional geometric configuration in the folding stable state more closely resembles an Archimedes' helix rather than a circle. Thus, in order to accurately and efficiently predict the folding stable state of the Bi-DCB, more accurate geometric equations describing the Bi-DCB in the folding stable state were established in this paper. Based on the energy principle, an analytical model was presented to predict the geometric configuration of the cross-section of the Bi-DCB in the folding stable state. In addition, based on the analytical model, Tsai-Hill and maximum stress criteria were utilized to analyze the failure index of the Bi-DCB in the folding stable state.

This paper is organized as follows: the geometrical model of the Bi-DCB is established in Section 2; an analytical model for predicting the geometric configuration of the cross-section of the Bi-DCB in the folding stable state is presented in Section 3; The stress analysis is conducted in Section 4; in Section 5, experiments and numerical simulation results are used to validate the analytical models presented in Sections 3 and 4; the key findings are summarized and conclusions are drawn in Section 6.

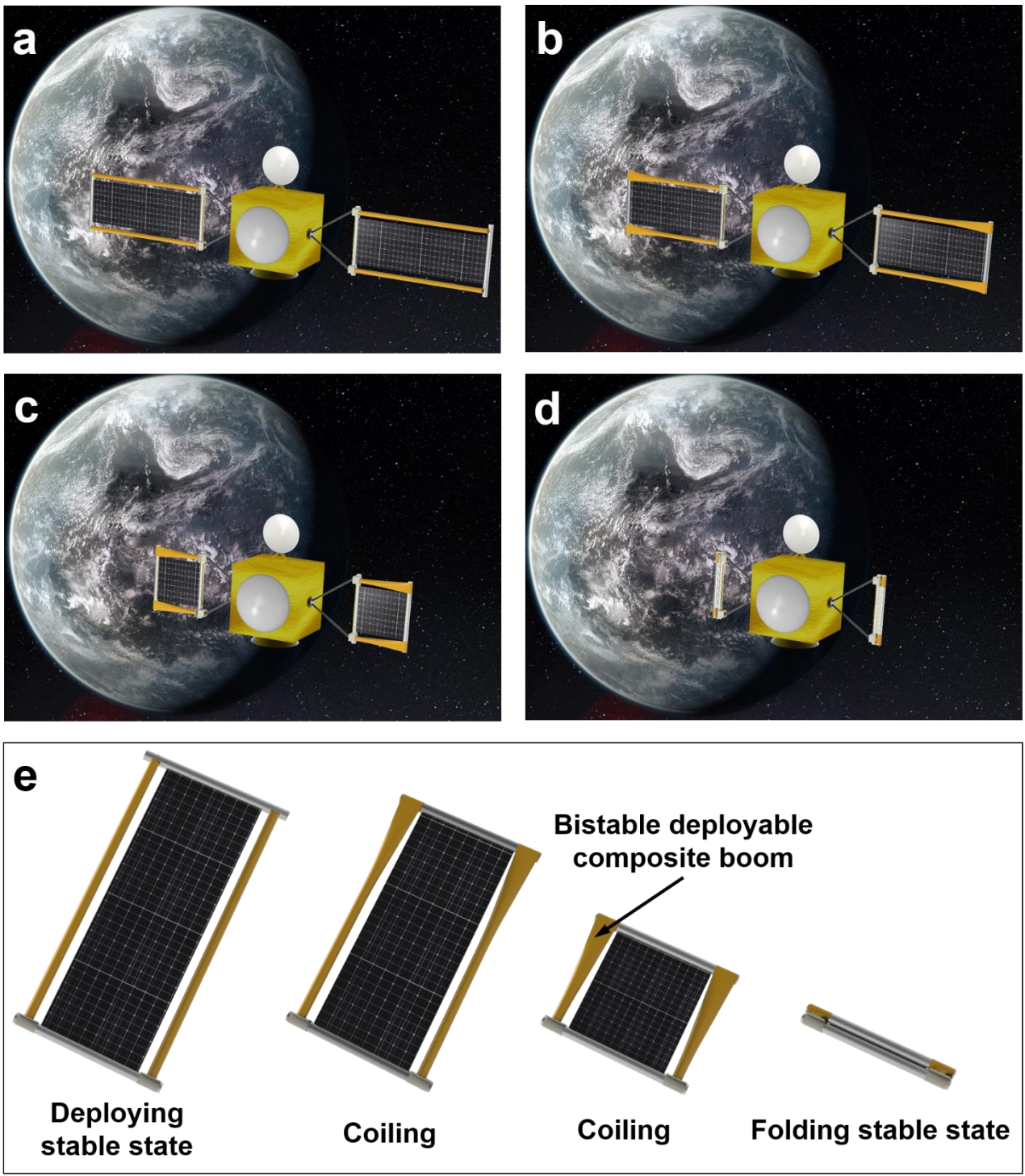

Fig. 1 Schematic diagrams of typical application and folding deformation process (a) Deploying stable state of the deployable solar array (b) Coiling process of the deployable solar array (c) Coiling process of the deployable solar array (d) Folding stable state of the deployable solar array (e) Bistable deformation processes of the deployable solar array and the Bi-DCB.

## 2. Geometrical model

The Bi-DCB can realize the conversion between the deploying stable state and the folding stable state by storing and releasing elastic strain energy, as shown in Fig. 2a. The curvature directions corresponding to the two stable states of the Bi-DCB are on the same side. A stable state is the initial deploying state, which has good bearing capacity; the other stable state is the folding state, where the Bi-DCB is closely and stably coiled together. The geometrical configuration of the Bi-DCB in the deploying stable state is determined by the length $L$, thickness $t$, radius $R$ and central angle $\theta$, as shown in Fig. 2b. To characterize the geometric behaviour of the Bi-DCB in the two stable states, the following basic assumptions are made in this paper:

(1) The central line of the longitudinal section of the Bi-DCB in the folding stable state is Archimedes' helix, and any two circles are closely attached (shown in Fig. 2c), which models the contact behavior between surfaces.

(2) The thickness change of the Bi-DCB shell is negligibly small in the bistable deformation process, so the overall deformation can be described with the changes of shape and curvature of the neutral surface without local extension.

(3) The curvature in the $y$ direction of the Bi-DCB is uniform in the bistable deformation process, and the curvature in the $y$ direction is zero in the folding stable state.

According to assumption (1) and Fig. 2c, the geometric configuration of the cross-section of the Bi-DCB in the folding stable state can be described by the multinomial shape function in a polar coordinate system of $(\rho,\alpha)$, that is

$$\rho = a\alpha + b, \quad (\alpha \in [\alpha_0, \alpha_1]) \tag{1}$$

where $a$ controls the distance between two adjacent circles, $b$ is the distance from the start-point to the origin of the polar coordinate system, and $\alpha_0$ and $\alpha_1$ are the polar angles at the start-point and the end-point of the Bi-DCB in the folding stable state, respectively.

According to assumption (1), when the Bi-DCB is in the folding stable state, any two circles are closely attached. It is possible to have

$$a = \frac{t}{2\pi} \tag{2}$$

According to geometric relationship, Eq. (1) should satisfy geometrical boundary condition as [21]

$$\begin{cases} a\alpha_0 + b = r_0 \\ a\alpha_1 + b = r_1 \end{cases} \tag{3}$$

where $r_0$ and $r_1$ are the polar radii at the start-point and the end-point of the Bi-DCB in the folding stable state, respectively.

Generally, $\alpha_0$ is zero, (i.e., $\alpha_0 = 0$), then Eq. (3) can be written as

$$\begin{cases} b = r_0 \\ \alpha_1 = \dfrac{r_1 - r_0}{a} \end{cases} \tag{4}$$

Substituting Eq. (2) into Eq. (4), it is possible to obtain as

$$\begin{cases} b = r_0 \\ \alpha_1 = \dfrac{2\pi(r_1 - r_0)}{t} \end{cases} \tag{5}$$

According to assumption (2), it is possible to obtain as

$$\int_{\alpha_0}^{\alpha_1} \sqrt{\rho^2 + (\rho')^2}\, d\alpha = L \tag{6}$$

where $L$ is the length of the Bi-DCB.

Substituting Eqs. (1) to (5) into Eq. (6) and integrating shows

$$\frac{t}{2\pi}\left[\frac{\pi r_1}{t}\sqrt{(\frac{2\pi r_1}{t})^2+1}+\frac{1}{2}\ln\left(\frac{\pi r_1}{t}+\sqrt{(\frac{2\pi r_1}{t})^2+1}\right)\right]-\frac{t}{2\pi}\left[\frac{\pi r_0}{t}\sqrt{(\frac{2\pi r_0}{t})^2+1}+\frac{1}{2}\ln\left(\frac{\pi r_0}{t}+\sqrt{(\frac{2\pi r_0}{t})^2+1}\right)\right]=L$$

(7)

Obviously, Eq. (7) is an implicit function, and the quantitative relationship between $r_0$ and $r_1$ can be determined by Newton iterative method. According to the definition of curvature, the curvature of the Bi-DCB in the $x$ direction (i.e., length direction) can be deduced as

$$\kappa_x=\frac{2(\rho')^2+\rho^2}{\left[(\rho')^2+\rho^2\right]^{\frac{3}{2}}}=\frac{\frac{t^2}{2\pi^2}+\rho^2}{\left(\frac{t^2}{4\pi^2}+\rho^2\right)^{\frac{3}{2}}},\quad \alpha\in[0,\frac{2\pi(r_1-r_0)}{t}]$$

(8)

## 3. Analytical model

The energy method is used to determine the geometric configuration of the Bi-DCB in the folding stable state. To establish the strain energy of the Bi-DCB in the folding stable state, the bistable deformation process is divided into two stages, as shown in Fig. 2d. Firstly, a bending moment $M_y$ is applied at the edge to change the curvature in the $y$ direction of the Bi-DCB from $1/R$ to 0. In the first stage, the external work is converted into bending strain energy. Then, a bending moment $M_x$ is applied at the edge to change the curvature in the $x$ direction of the Bi-DCB from 0 to $\kappa_x$. Similarly, the external work is converted into bending strain energy in the second stage.

The applicability of classical laminate theory for predicting the folding stable state of the Bi-DCB has been demonstrated in previous studies, including four classical analytical models [17-20]. Similarly, utilizing an ABD matrix of the classical laminate theory, the relationship between internal forces and deformation of the Bi-DCB with antisymmetric layup in bistable deformation process can be expressed as

$$\begin{bmatrix} N_x \\ N_y \\ N_{xy} \\ M_x \\ M_y \\ M_{xy} \end{bmatrix} = \begin{bmatrix} A_{11} & A_{12} & 0 & 0 & 0 & B_{16} \\ A_{12} & A_{22} & 0 & 0 & 0 & B_{26} \\ 0 & 0 & A_{66} & B_{16} & B_{26} & 0 \\ 0 & 0 & B_{16} & D_{11} & D_{12} & 0 \\ 0 & 0 & B_{26} & D_{12} & D_{22} & 0 \\ B_{16} & B_{26} & 0 & 0 & 0 & D_{66} \end{bmatrix} \begin{bmatrix} \Delta\varepsilon_x \\ \Delta\varepsilon_y \\ \Delta\gamma_{xy} \\ \Delta\kappa_x \\ \Delta\kappa_y \\ \Delta\kappa_{xy} \end{bmatrix} \quad (9)$$

where

$$\begin{cases} N_x = 0 \\ N_y = 0 \\ N_{xy} = 0 \\ M_{xy} = 0 \\ \Delta\kappa_{xy} = 0 \end{cases} \quad (10)$$

Mansfield deduced the bending strain energy per unit area in the mid-surface of a laminated plate [22], as follows:

$$u = \frac{1}{2}(M_x \Delta\kappa_x + M_y \Delta\kappa_y + M_{xy} \Delta\kappa_{xy}) \quad (11)$$

where

$$\begin{cases} \Delta\kappa_x = \kappa_x \\ \Delta\kappa_y = \kappa_y - \dfrac{1}{R} \end{cases} \quad (12)$$

Substituting Eqs. (9), (10) and (12) into Eq. (11), it is possible to deduce as

$$u = \frac{1}{2}\kappa_x\left[B_{16}\gamma_{xy} + D_{11}\kappa_x + D_{12}(\kappa_y - \frac{1}{R})\right] + \frac{1}{2}(\kappa_y - \frac{1}{R})\left[B_{26}\gamma_{xy} + D_{12}\kappa_x + D_{22}(\kappa_y - \frac{1}{R})\right] \quad (13)$$

Substituting Eq. (10) into Eq. (9) and solving Eq. (9), it is possible to determine as

$$\gamma_{xy} = B^*_{61}\kappa_x + B^*_{62}\kappa_y \quad (14)$$

where

$$\boldsymbol{B}^* = -\boldsymbol{A}^{-1}\boldsymbol{B} \quad (15)$$

According to assumption (3), $u$ is uniform. Substituting Eqs. (14) and (15) into Eq. (13), the expression of the strain energy for a unit longitudinal length of Bi-DCB in the folding stable state is given as

$$U = \frac{R\theta}{2}\left[(D_{11}+B_{16}B_{61}^*)\kappa_x^2 + (2D_{12}+B_{16}B_{62}^*+B_{26}B_{61}^*)(\kappa_y-\frac{1}{R})\kappa_x + (D_{22}+B_{26}B_{62}^*)(\kappa_y-\frac{1}{R})^2\right] \quad (16)$$

where $R$ and $\theta$ are the radius and central angle of the cross-section of the Bi-DCB, respectively.

According to assumption (3), when the Bi-DCB is in the folding stable state, the curvature in the $y$ direction is 0 (i.e., $\kappa_y = 0$). Eq. (16) can be simplified to

$$U = \frac{R\theta}{2}\left[(D_{11}+B_{16}B_{61}^*)\kappa_x^2 - (2D_{12}+B_{16}B_{62}^*+B_{26}B_{61}^*)\frac{\kappa_x}{R} + (D_{22}+B_{26}B_{62}^*)\frac{1}{R^2}\right] \quad (17)$$

Therefore, the total strain energy of the Bi-DCB in the folding stable state can be expressed as

$$\Pi = \int_{\alpha_0}^{\alpha_1} U(\alpha)\rho(\alpha)d\alpha \quad (18)$$

Substituting Eqs. (8) and (17) into Eq. (18), it is possible to obtain as

$$\Pi = \frac{R\theta}{2}\int_0^{\frac{2\pi(r_1-r_0)}{t}} \left\{ (D_{11}+B_{16}B_{61}^*)\frac{\left[\frac{t^2}{2\pi^2}+(\frac{t\alpha}{2\pi}+r_0)^2\right]^2}{\left[\frac{t^2}{4\pi^2}+(\frac{t\alpha}{2\pi}+r_0)^2\right]^3} + (D_{22}+B_{26}B_{62}^*)\frac{1}{R^2} \right. \\ \left. -(2D_{12}+B_{16}B_{62}^*+B_{26}B_{61}^*)\frac{\frac{t^2}{2\pi^2}+(\frac{t\alpha}{2\pi}+r_0)^2}{R[\frac{t^2}{4\pi^2}+(\frac{t\alpha}{2\pi}+r_0)^2]^{\frac{3}{2}}} \right\} (\frac{t\alpha}{2\pi}+r_0)d\alpha \quad (19)$$

$\alpha$ in Eq. (19) is represented by $\rho$, and Eq. (19) can be simplified as

$$\Pi = \frac{\pi R\theta}{t}\int_{r_0}^{r_1}\left[(D_{11}+B_{16}B_{61}^*)\frac{(\frac{t^2}{2\pi^2}+\rho^2)^2}{(\frac{t^2}{4\pi^2}+\rho^2)^3} + (D_{22}+B_{26}B_{62}^*)\frac{1}{R^2} \right. \\ \left. -(2D_{12}+B_{16}B_{62}^*+B_{26}B_{61}^*)\frac{\frac{t^2}{2\pi^2}+\rho^2}{R(\frac{t^2}{4\pi^2}+\rho^2)^{\frac{3}{2}}}\right]\rho d\rho \quad (20)$$

It is known that the folding stable state of the Bi-DCB corresponds to an energy minimum. Therefore, the minimum energy principle can be used for Eq. (20). It is possible to have

$$\frac{d\Pi}{dr_0} = r_1 H_1 \frac{dr_1}{dr_0} - r_0 H_2 = 0 \tag{21}$$

where $H_1$ and $H_2$ are transformation variables.

$$H_1 = \frac{\pi R\theta}{t} \left[ \begin{array}{c} (D_{11}+B_{16}B_{61}^*)\dfrac{(\dfrac{t^2}{2\pi^2}+r_1^2)^2}{(\dfrac{t^2}{4\pi^2}+r_1^2)^3} + (D_{22}+B_{26}B_{62}^*)\dfrac{1}{R^2} \\ \\ -(2D_{12}+B_{16}B_{62}^*+B_{26}B_{61}^*)\dfrac{\dfrac{t^2}{2\pi^2}+r_1^2}{R(\dfrac{t^2}{4\pi^2}+r_1^2)^{\frac{3}{2}}} \end{array} \right] \tag{22}$$

$$H_2 = \frac{\pi R\theta}{t} \left[ \begin{array}{c} (D_{11}+B_{16}B_{61}^*)\dfrac{(\dfrac{t^2}{2\pi^2}+r_0^2)^2}{(\dfrac{t^2}{4\pi^2}+r_0^2)^3} + (D_{22}+B_{26}B_{62}^*)\dfrac{1}{R^2} \\ \\ -(2D_{12}+B_{16}B_{62}^*+B_{26}B_{61}^*)\dfrac{\dfrac{t^2}{2\pi^2}+r_0^2}{R(\dfrac{t^2}{4\pi^2}+r_0^2)^{\frac{3}{2}}} \end{array} \right] \tag{23}$$

It should be noted that there is a differential term $\dfrac{dr_1}{dr_0}$ in Eq. (21), and the central difference algorithm is used to achieve numerical differentiation.

Obviously, Eqs. (7) and (21) contain only two unknown variables, namely $r_0$ and $r_1$. The two unknown variables can be obtained by combining Eqs. (7) and (21) and solving them numerically. Substituting the obtained $r_0$ and $r_1$ into Eq. (5), it is possible to obtain the polar angle $\alpha_1$.

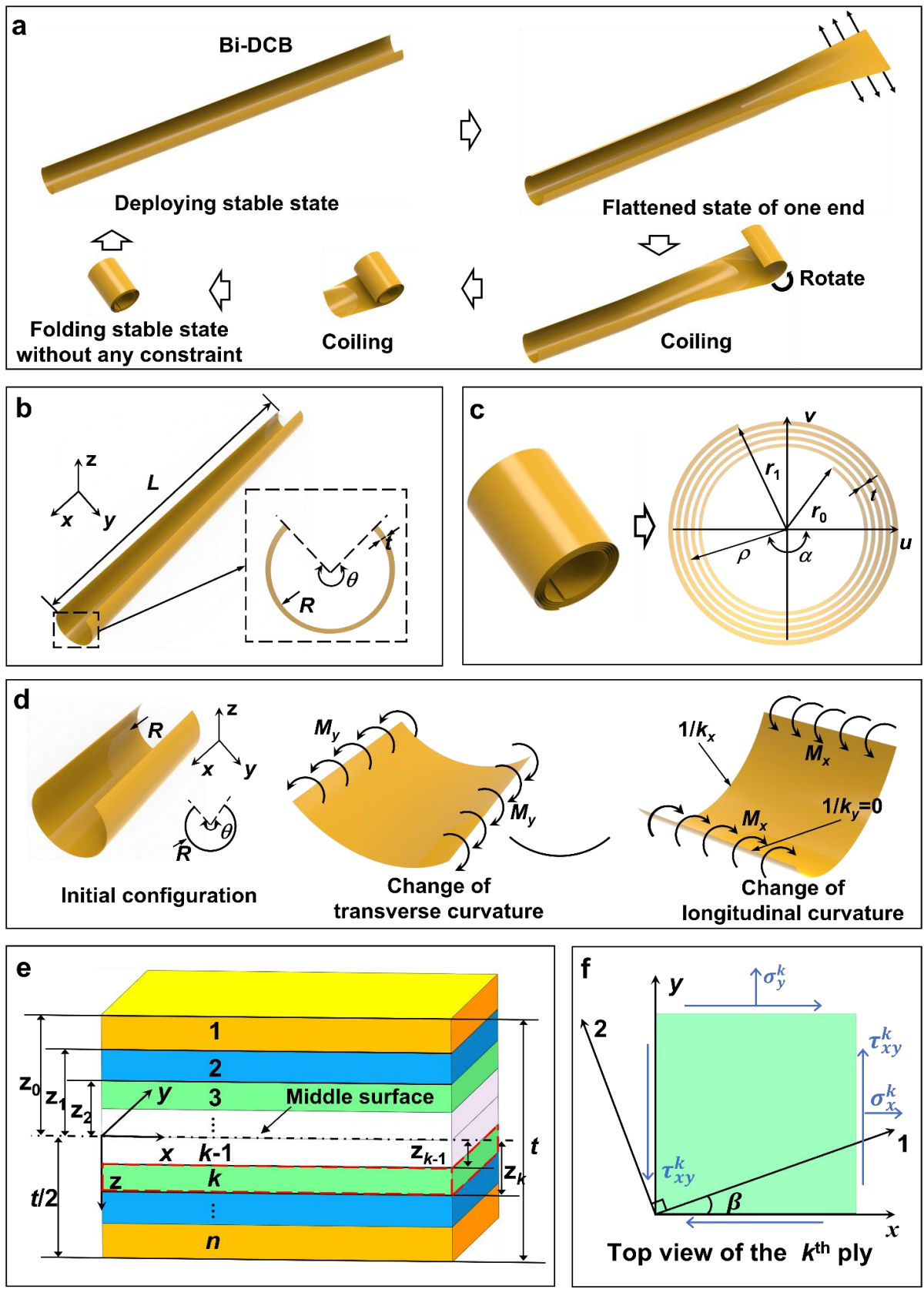

Fig. 2 Analytical model for the Bi-DCB: (a) Bistable deformation process (b) Deploying stable state (c) Folding stable state (d) Force analysis and deformation (e) Schematic diagram of the laminate ply (f) Stress analysis.

## 4. Stress analysis

In the bistable deformation process of the Bi-DCB, analyzing the stress level in the folding stable state is crucial. Stress analysis ensures stability and reliability in practical applications. Understanding the stress distribution in the folding stable state can help optimize the design and improve structural performance. In addition, failure risks may be overlooked if stress levels in the folding stable state are not analyzed. Moreover, the stress distribution in the folding stable state can also affect the deploying process of the Bi-DCB, so it is important to analyze it to avoid jamming or deformation during the deploying process.

According to assumption (1), the Bi-DCB has the maximum strain and stress in the *x* direction at the start-point when it is in the folding stable state. The maximum value of the change of curvature in the *x* direction is

$$\Delta \kappa_x \big|_{\alpha=0} = \frac{\dfrac{t^2}{2\pi^2} + r_0^2}{(\dfrac{t^2}{4\pi^2} + r_0^2)^{\frac{3}{2}}} \tag{24}$$

The maximum value of the change of curvature in the *y* direction is

$$\Delta \kappa_y \big|_{\alpha=0} = -\frac{1}{R} \tag{25}$$

Fig. 2e is the schematic diagram of a laminate. The stress-strain relationship of the $k^{\text{th}}$ ply ($k=1, 2,3...n$) at the dangerous point is

$$\begin{bmatrix} \sigma_x \\ \sigma_y \\ \tau_{xy} \end{bmatrix}_k = \begin{bmatrix} \overline{Q}_{11} & \overline{Q}_{12} & \overline{Q}_{16} \\ \overline{Q}_{12} & \overline{Q}_{22} & \overline{Q}_{26} \\ \overline{Q}_{16} & \overline{Q}_{26} & \overline{Q}_{66} \end{bmatrix}_k \begin{bmatrix} z\Delta\kappa_x \\ z\Delta\kappa_y \\ z\Delta\kappa_{xy} \end{bmatrix} \tag{26}$$

Substituting Eqs. (24) and (25) into Eq. (26), it is possible to determine as

$$\sigma_x^k = z\overline{Q}_{11}^k \Delta\kappa_x + z\overline{Q}_{12}^k \Delta\kappa_y \tag{27}$$

$$\sigma_y^k = z\overline{Q}_{12}^k \Delta\kappa_x + z\overline{Q}_{22}^k \Delta\kappa_y \tag{28}$$

$$\tau_{xy}^{k} = z\overline{Q}_{16}^{k}\Delta\kappa_{x} + z\overline{Q}_{26}^{k}\Delta\kappa_{y} \tag{29}$$

According to the coordinate transformation equation of stress components of the $k^{th}$ ply in the laminate (shown in Fig. 2f), the maximum principal stresses of the $k^{th}$ ply in the laminate can be expressed as

$$\begin{bmatrix} \sigma_{1} \\ \sigma_{2} \\ \tau_{12} \end{bmatrix}_{k} = \begin{bmatrix} \cos^{2}\beta & \sin^{2}\beta & 2\sin\beta\cos\beta \\ \sin^{2}\beta & \cos^{2}\beta & -2\sin\beta\cos\beta \\ -\sin\beta\cos\beta & \sin\beta\cos\beta & \cos^{2}\beta-\sin^{2}\beta \end{bmatrix}_{k} \begin{bmatrix} \sigma_{x} \\ \sigma_{y} \\ \tau_{xy} \end{bmatrix}_{k} \tag{30}$$

Substituting Eqs. (27) to (29) into Eq. (30), it is possible to deduce as

$$\begin{aligned} \sigma_{1}^{k} &= z\cos^{2}\beta(\overline{Q}_{11}^{k}\Delta\kappa_{x} + \overline{Q}_{12}^{k}\Delta\kappa_{y}) + z\sin^{2}\beta(\overline{Q}_{12}^{k}\Delta\kappa_{x} + \overline{Q}_{22}^{k}\Delta\kappa_{y}) \\ &+ 2z\sin\beta\cos\beta(\overline{Q}_{16}^{k}\Delta\kappa_{x} + \overline{Q}_{26}^{k}\Delta\kappa_{y}) \end{aligned} \tag{31}$$

$$\begin{aligned} \sigma_{2}^{k} &= \sin^{2}\beta(\overline{Q}_{11}^{k}\Delta\kappa_{x} + \overline{Q}_{12}^{k}\Delta\kappa_{y}) + \cos^{2}\beta(\overline{Q}_{12}^{k}\Delta\kappa_{x} + \overline{Q}_{22}^{k}\Delta\kappa_{y}) \\ &- 2\sin\beta\cos\beta(\overline{Q}_{16}^{k}\Delta\kappa_{x} + \overline{Q}_{26}^{k}\Delta\kappa_{y}) \end{aligned} \tag{32}$$

$$\begin{aligned} \tau_{12}^{k} &= -z\sin\beta\cos\beta(\overline{Q}_{11}^{k}\Delta\kappa_{x} + \overline{Q}_{12}^{k}\Delta\kappa_{y}) + z\sin\beta\cos\beta(\overline{Q}_{12}^{k}\Delta\kappa_{x} + \overline{Q}_{22}^{k}\Delta\kappa_{y}) \\ &+ z(\cos^{2}\beta - \sin^{2}\beta)(\overline{Q}_{16}^{k}\Delta\kappa_{x} + \overline{Q}_{26}^{k}\Delta\kappa_{y}) \end{aligned} \tag{33}$$

The Tsai-Hill criterion and maximum stress criterion can effectively determine the stress level of deployable composite structures during large deformation [23-31]. Based on the above two failure criteria, the failure index is calculated to analyze the stress level of the Bi-DCB in the folding stable state. When the failure index reaches or exceeds 1, the Bi-DCB fails; otherwise, it does not fail.

The Tsai-Hill failure index $I_{f,1}$ can be expressed as [25]

$$I_{f,1}^{2} = \frac{(\sigma_{1}^{k})^{2}}{X_{1}^{2}} - \frac{\sigma_{1}^{k}\sigma_{2}^{k}}{X_{2}^{2}} + \frac{(\sigma_{2}^{k})^{2}}{Y^{2}} + \frac{(\tau_{12}^{k})^{2}}{S_{12}^{2}} \tag{34}$$

where

$$X_{1} = \begin{cases} X_{t} & \text{if } \sigma_{1}^{k} > 0 \\ X_{c} & \text{if } \sigma_{1}^{k} < 0 \end{cases} \tag{35}$$

$$X_2 = \begin{cases} X_t & \text{if } \sigma_2^k > 0 \\ X_c & \text{if } \sigma_2^k < 0 \end{cases} \quad (36)$$

$$Y = \begin{cases} Y_t & \text{if } \sigma_2^k > 0 \\ Y_c & \text{if } \sigma_2^k < 0 \end{cases} \quad (37)$$

where $\sigma_1^k$ and $\sigma_2^k$ are the stresses of the $k^{th}$ ply in longitudinal and transverse direction respectively, $\tau_{12}^k$ is the shear stress of the $k^{th}$ ply in longitudinal and transverse direction, $X_1$ and $X_2$ are the longitudinal strength of composite ply, $X_t$ and $X_c$ are longitudinal tensile and compressive strength of composite ply, $Y$ is the transverse strength of composite ply, $Y_t$ and $Y_c$ are transverse tensile and compressive strength of composite ply, and $S_{12}$ is the in-plane shear strength of composite ply.

The maximum stress failure index $I_{f,2}$ can be expressed as [25]

$$I_{f,2} = \max \left\{ \left| \frac{\sigma_1^k}{X_1} \right|, \left| \frac{\sigma_2^k}{Y} \right|, \frac{\tau_{12}^k}{S_{12}} \right\} \quad (38)$$

## 5. Results and discussion

### 5.1 Analytical model validation

Based on the analytical model for predicting the folding stable state of the Bi-DCB established in Sections 2 to 4, the solution procedure was designed with MATLAB software. To validate the analytical model proposed in this research, the prediction results are compared with experimental results and FEM results in literature [32]. A total of six Bi-DCB specimens were prepared. Table 1 lists basic properties of T700/epoxy ply, and Table 2 lists geometric parameters and stacking sequence of six Bi-DCB specimens. For experiment, six specimens were manually coiled into folding stable states. A camera was used to capture the cross-sectional images of six specimens in the folding stable state to obtain the geometric configuration. The images were imported into GetData software to extract the coordinates of 30 data points on the neutral surface. The coordinates were then linearly fitted with Archimedes' helix, and the polar radii and angle were recorded at the starting and end points. In addition, a three-dimensional geometric nonlinear explicit FEM 1 was established. In the step 1, one end of the Bi-DCB was stretched until it was fully flattened. The step 2 was used to simulate the coiling process of the Bi-DCB along the roller. In the step 3, the constraints and all boundary conditions were removed to allow the Bi-DCB to reach the folding stable state. In the FEM 1, the self- contact and the contact between the Bi-DCB and roller were set as the general contact. Moreover, a three-dimensional geometric nonlinear explicit FEM 2 was established. Although the predicted results from both FEMs were almost identical, the modeling approaches were different. In the FEM 2, a rigid plate was utilized to assist in the coiling of the Bi-DCB. The reason for establishing two FEMs in reference [32] was to provide readers with more modeling options, facilitating the replication of the bistable deformation process. For further details on the experiments, FEM 1, and FEM 2, please refer to reference [32].

Fig. 3 shows total strain energy versus start-point polar radius curves of the six Bi-DCBs. It is shown that the folding stable state of the Bi-DCB is related to the existence of the local minimum of total strain energy. Fig. 4 illustrates the geometric configuration of the Bi-DCB in the folding stable state determined by experiments, two FEMs and analytical model, and the four are in good agreement. Archimedes' helix was used to linearly fit the data points of the cross-section of the folding stable state determined by experiments, two FEMs and analytical model. The fitted Archimedes' helix is shown in Fig. 5, and the polar radii $r_0$ and $r_1$ at the start-point and the end-point as well as the polar angle $\alpha_1$ of Archimedes' helix are listed in Table 3. It is clear that the geometric configuration of the folding stable state, the polar radii $r_0$ and $r_1$ at the start-point and the end-point as well as the polar angle $\alpha_1$ of Archimedes' helix are in good agreement. For the six Bi-DCBs, the maximum errors between the polar radii $r_0$ and $r_1$ at the start-point and the end-point as well as the polar angle $\alpha_1$ of Archimedes' helix predicted by the analytical model and experiments are 9.42%, 8.82% and 10.08% respectively; the maximum errors between the polar radii at the start-point and the end-point as well as the polar angle of Archimedes' helix predicted by the analytical model and the FEM 1 results are 7.21%, 7.67% and 6.93% respectively; the maximum errors between the polar radii at the start-point and the end-point as well as the polar angle of Archimedes' helix predicted by the analytical model and the FEM 2 results are 6.85%, 7.48% and 6.69% respectively.

The precision of the analytical model developed in this study is sufficient for engineering applications. Although the maximum error of the results is 10.08%, errors within 10% are generally acceptable and usable in many engineering applications. The focus of the analytical model is to provide insights into the folding stable state of the Bi-DCB, rather than to determine final parameters with extreme precision. Due to the complexity of bistable structures and inherent experimental uncertainties, an error of around 10% is inevitable during the initial design stage. The purpose of the analytical model is to achieve a basic understanding and preliminary optimization, and to guide more refined simulations and experiments to improve accuracy. Potential sources of errors mainly include: (1) finite element modeling, which is a discretization process resulting in predicted stiffness being greater than actual stiffness; (2) analytical modeling, which simplifies geometric configuration; and (3) the complexity of bistable structures and experimental uncertainties.

The two failure criteria in Section 4 (namely, Tsai-Hill criterion and maximum stress criterion) are applied to calculate the maximum failure index of each layer of six Bi-DCBs in the folding stable state, and the maximum failure indices are listed in Fig. 6 and Table 4. According to Fig. 6 and Table 4, among all the failure indices, the maximum values of FEM 1 and FEM 2 are 0.3694 and 0.3705 respectively, and the maximum value of the analytical model is 0.3445. It means that the six Bi-DCBs can realize the bistable function without failure, which is consistent with experimental results. In conclusion, the analytical model proposed in this research shows good prediction accuracy for evaluating the folding stable state of the Bi-DCB, which proves the validity of the analytical model.

Table 1  Specifications and properties of T700/epoxy composite ply [32].

| Specifications and properties | Values |
|---|---|
| Longitudinal elastic modulus $E_1$ (GPa) | 128.61 |
| Transverse elastic modulus $E_2$ (GPa) | 7.52 |
| In-plane shear modulus $G_{12}$ (GPa) | 4.82 |
| Inter-laminar shear modulus $G_{13}$ (GPa) | 4.50 |
| Inter-laminar shear modulus $G_{23}$ (GPa) | 4.50 |
| Poisson's ratio $v_{12}$ | 0.314 |
| Poisson's ratio $v_{21}$ | 0.018 |
| Longitudinal tensile strength $X_t$ (MPa) | 2103.44 |
| Transverse tensile strength $Y_t$ (MPa) | 75.97 |
| Longitudinal compressive strength $X_c$ (MPa) | 1233.65 |
| Transverse compressive strength $Y_c$ (MPa) | 181.46 |
| In-plane shear strength $S_{12}$ (MPa) | 216.36 |
| Density (g/mm$^3$) | 1.60×10$^{-3}$ |
| Ply thickness (mm) | 0.03 |

Table 2  Specifications of the six Bi-DCB specimens.

| No. | Radius (mm) | Arc length (mm) | Center angle (deg) | Length (mm) | Stacking Configuration |
|---|---|---|---|---|---|
| 1 | 20 | 80 | 229 | 620 | [45/-45/45/-45/45/90/-45/45/-45/45/-45] |
| 2 | 20 | 80 | 229 | 400 | [45/-45/45/-45/0/45/-45/45/-45] |
| 3 | 20 | 80 | 229 | 200 | [45/-45/45/-45/90/45/-45/45/-45] |
| 4 | 27.5 | 80 | 167 | 650 | [45/-45/45/-45/0/90/45/-45/45/-45] |
| 5 | 27.5 | 80 | 167 | 650 | [45/-45/45/-45/45/-45/45/-45/45/-45] |
| 6 | 30 | 100 | 191 | 620 | [45/-45/45/-45/45/90/-45/45/-45/45/-45] |

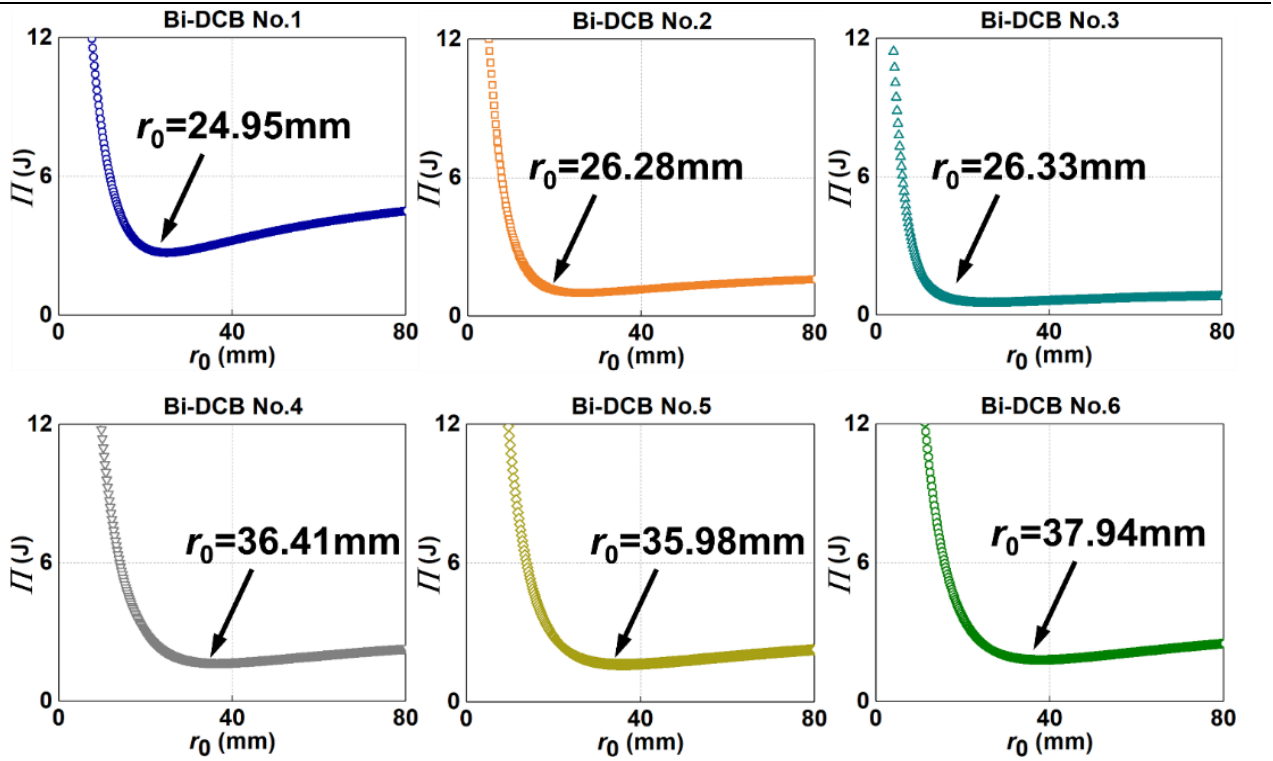

Fig. 3 Total strain energy versus polar radius at the start-point curves of six Bi-DCBs.

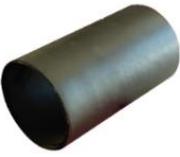

Fig. 4 Geometric configuration of six Bi-DCBs in the folding stable state.

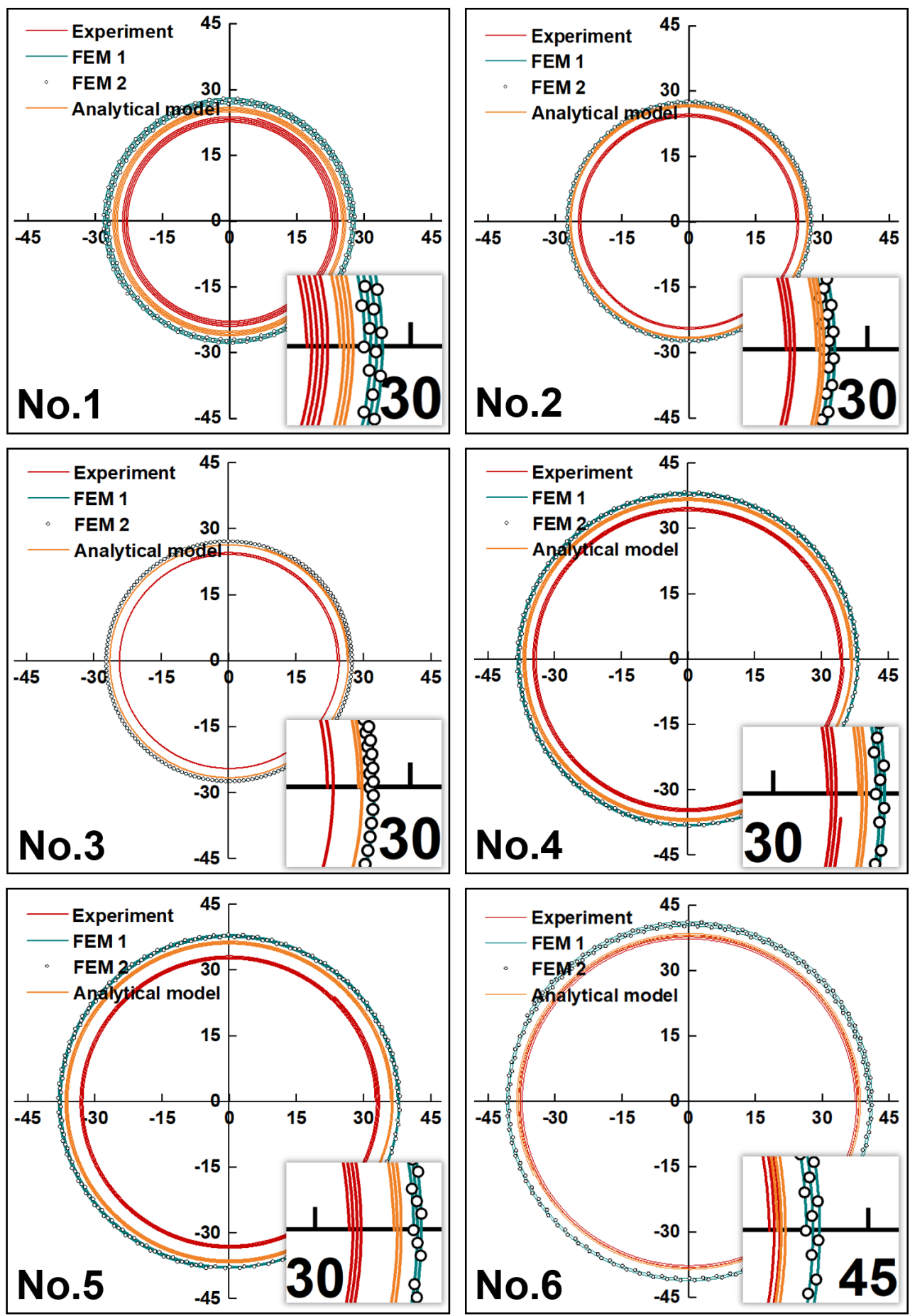

Fig. 5 Geometric configuration of the cross-section of six Bi-DCBs in the folding stable state.

Table 3  Polar radii at the start-point and the end-point as well as polar angle of six Bi-DCBs in the folding stable state.

| Geometrical parameter | | | No.1 | No.2 | No.3 | No.4 | No.5 | No.6 |
|---|---|---|---|---|---|---|---|---|
| $r_0$ (mm) | Analytical model | | 24.95 | 26.28 | 26.33 | 36.41 | 35.98 | 37.94 |
| | Experiment | Value | 22.63 | 24.04 | 24.13 | 34.08 | 32.59 | 37.31 |
| | | Error | 9.30% | 8.52% | 8.36% | 6.40% | 9.42% | 1.66% |
| | FEM 1 | Value | 26.75 | 26.89 | 27.16 | 37.75 | 37.49 | 40.21 |
| | | Error | 7.21% | 2.32% | 3.15% | 3.68% | 4.20% | 5.98% |
| | FEM 2 | Value | 26.66 | 26.85 | 27.15 | 37.68 | 37.47 | 40.15 |
| | | Error | 6.85% | 2.17% | 3.11% | 3.49% | 4.14% | 5.82% |
| $r_1$ (mm) | Analytical model | | 26.22 | 26.93 | 26.65 | 37.25 | 36.83 | 38.79 |
| | Experiment | Value | 24.23 | 24.87 | 24.66 | 35.08 | 33.58 | 38.41 |
| | | Error | 7.59% | 7.65% | 7.47% | 5.82% | 8.82% | 0.98% |
| | FEM 1 | Value | 28.23 | 27.75 | 27.52 | 38.57 | 38.32 | 41.47 |
| | | Error | 7.67% | 3.04% | 3.26% | 3.54% | 4.04% | 6.91% |
| | FEM 2 | Value | 28.18 | 27.67 | 27.44 | 38.55 | 38.31 | 41.35 |
| | | Error | 7.48% | 2.75% | 2.94% | 3.49% | 4.02% | 6.60% |
| $\alpha_1$ (rad) | Analytical model | | 24.23 | 15.04 | 7.55 | 17.65 | 17.85 | 16.16 |
| | Experiment | Value | 26.46 | 16.36 | 8.20 | 18.80 | 19.65 | 16.38 |
| | | Error | 9.20% | 8.78% | 8.61% | 6.52% | 10.08% | 1.36% |
| | FEM 1 | Value | 22.55 | 14.64 | 7.32 | 17.03 | 17.15 | 15.07 |
| | | Error | 6.93% | 2.66% | 3.05% | 3.51% | 3.92% | 6.75% |
| | FEM 2 | Value | 22.61 | 14.67 | 7.33 | 17.05 | 17.15 | 15.11 |
| | | Error | 6.69% | 2.46% | 2.91% | 3.40% | 3.92% | 6.50% |

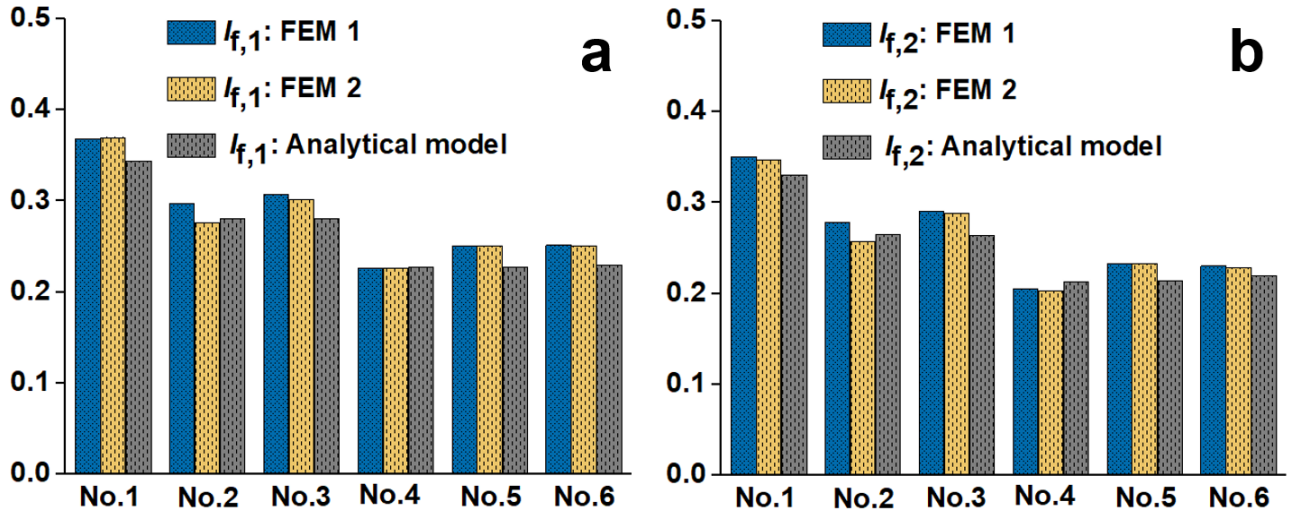

Fig. 6 Comparison of the failure indices of six Bi-DCBs in the folding stable state: (a) Tsai-Hill failure index (b) Maximum stress failure index.

Table 4 Failure indices of six Bi-DCBs in the folding stable state.

| Failure index | | | No.1 | No.2 | No.3 | No.4 | No.5 | No.6 |
|---|---|---|---|---|---|---|---|---|
| $I_{f,1}$ | Analytical model | | 0.3445 | 0.2810 | 0.2810 | 0.2270 | 0.2271 | 0.2294 |
| | FEM 1 | Value | 0.3694 | 0.2981 | 0.3079 | 0.2266 | 0.2505 | 0.2520 |
| | | Error | 7.23% | 6.09% | 9.57% | 0.18% | 10.30% | 9.85% |
| | FEM 2 | Value | 0.3705 | 0.2769 | 0.3024 | 0.2270 | 0.2509 | 0.2509 |
| | | Error | 7.55% | 1.46% | 7.62% | 0% | 10.48% | 9.37% |
| $I_{f,2}$ | Analytical model | | 0.3311 | 0.2648 | 0.2646 | 0.2133 | 0.2144 | 0.2194 |
| | FEM 1 | Value | 0.3514 | 0.2786 | 0.2907 | 0.2058 | 0.2333 | 0.2304 |
| | | Error | 6.13% | 5.21% | 9.86% | 3.52% | 8.81% | 5.01% |
| | FEM 2 | Value | 0.3477 | 0.2577 | 0.2880 | 0.2034 | 0.2328 | 0.2284 |
| | | Error | 5.01% | 2.68% | 8.84% | 4.64% | 8.58% | 4.10% |

**5.2 Engineering application significance**

Despite the existence of several analytical models in previous studies (e.g., the classical two-parameter model) [17-20], it is suggested that these models may not be entirely appropriate for predicting the folding stable state of Bi-DCBs, regardless of their length. This is due to the assumption in these models that the cross-section of the folding stable state is circular, which limits its applicability to very short bistable cylindrical shells with an arc-shaped folding state cross-section (i.e., less than one turn). However, even relatively shorter-length Bi-DCBs used in space applications typically have a folding stable state cross-section that is helical, not circular, and may have several turns. For longer-length Bi-DCBs, the analytical model established in this paper is more capable of revealing the true and complete geometric configuration of the folding stable state compared to the classical two-parameter model. The longest specimen length of the aforementioned Bi-DCB specimens is only 650 mm, while the boom length of solar sails for space applications can exceed 5000 mm, such as the NanoSail-D2 [33], LightSail 1 [34], LightSail 2 [35] and NEA Scout [36] satellites. Therefore, in order to demonstrate the advantages of the analytical model in this study, a 5000 mm long Bi-DCB, labelled as No.7, was designed in this section, with identical geometric parameters and stacking configuration as No.1.

The cross-sectional geometric configurations of designs No.1 and No.7 were predicted by using both the classical two-parameter analytical model and the analytical model in this study, as illustrated in Figs. 7a and 7b. Specifically, when the length reaches 5000 mm, there is a significant discrepancy in the predicted cross-sectional geometric configurations between the two models. The analytical model in this study yields a prediction result that aligns more closely with the actual ensemble configuration. According to Fig. 7c, the Tsai-Hill and maximum stress failure indices of design No.7 are larger than those of design No.1, since the start-point of the folding stable state of No.7 exhibits a larger curvature variation in the $x$ direction compared to that of No.1. Fig. 7d reveals that, in comparison to design No.1, the polar radii at both the start-point and end-point of the folding stable configuration of design No.7 exhibit a larger difference. This indicates that length plays a crucial role in affecting the cross-sectional geometric configuration of the folding stable state. As the length increases, the helical characteristic of the folding stable state becomes more pronounced.

In conclusion, the analytical model in this study exhibits broader applicability when compared to the classical two-parameter analytical model. The main contribution of this paper is to establish an analytical model that can predict the true and complete geometric configuration of the folding stable state of the Bi-DCB with sufficient predictive accuracy for engineering applications.

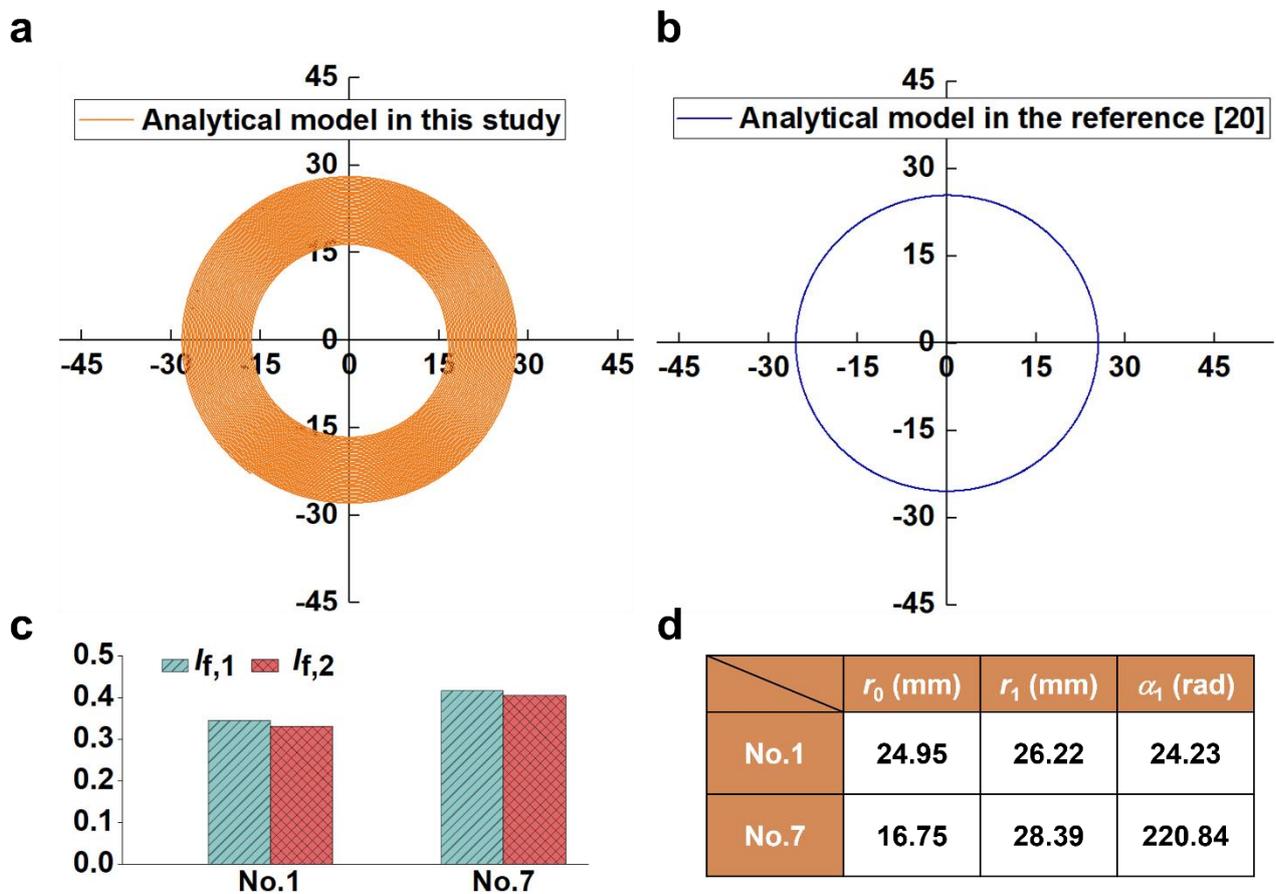

Fig. 7 Effect of length on the folding stable state of the Bi-DCB: (a) Cross-sectional geometric configuration predicted by the analytical model in this study (b) Cross-sectional geometric configuration predicted by the classical two-parameter analytical model in the reference [20] (c) Failure indices (d) Polar radii at both the start-point and end-point as well as the polar angle.

## 5.3 Parametric study

According to the analytical model, it is known that geometric parameters of the Bi-DCB in deploying stable state can significantly affect the folding stable state of the Bi-DCB, such as the radius of the cross-section, thickness and length. According to Fig. 6 and Table 3, the analytical model proposed in this paper and the two FEMs can accurately predict the folding stable state of the Bi-DCB, but compared with the two FEMs, the analytical model can save a lot of time. Therefore, the analytical model is used to explore the effect of the radius of the cross-section, thickness and length on the folding stable state of the Bi-DCB. The material properties and remaining geometric parameters are consistent with the Bi-DCB No.4. The specific variable values are as follows:

- the radius of the cross-section is varied as 22.5mm, 25mm, 27.5mm, 30mm and 32.5mm;
- the thickness is varied as 0.24mm, 0.27mm, 0.3mm, 0.33mm and 0.36mm;
- the length is varied as 250mm, 450mm, 650mm, 850mm and 1050mm.

It is clear from Figs. 8a to 8c that the local minimum of total strain energy exists in the Bi-DCBs with different radii of the cross-section. With the increase of the radius of the cross-section, the polar radii $r_0$ and $r_1$ at the start-point and the end-point of Archimedes' helix increase, while the polar angle $\alpha_1$ of Archimedes' helix decreases. In addition, the failure indices of the Bi-DCBs with different radii of the cross-section do not reach 1 when they are in the folding stable state, indicating that the Bi-DCBs can achieve the bistable function without failure. With the increase of the radius of the cross-section, the Tsai-Hill failure index and the maximum stress failure index decrease significantly. The reason is that the change of curvature in the $y$ direction in Eq. (25) decreases, resulting in the decrease of the Tsai-Hill failure index and the maximum stress failure index.

It is clear from Figs. 8d to 8f that the local minimum of total strain energy exists in the Bi-DCBs with different thicknesses. With the increase of the thickness, the polar radii $r_0$ and $r_1$ at the start-point and the end-point of Archimedes' helix increase, while the polar angle $α_1$ of Archimedes' helix decreases. In addition, the failure indices of the Bi-DCBs with different thicknesses do not reach 1 when they are in the folding stable state, indicating that the Bi-DCBs can achieve the bistable function without failure. With the increase of the thickness, the Tsai-Hill failure index and the maximum stress failure index increase. The reason is that the values of $z$ in Eqs. (26) to (33) increase, resulting in the Tsai-Hill failure index and the maximum stress failure index increase.

It is clear from Figs. 8g to 8i that the local minimum of total strain energy exists in the Bi-DCBs with different lengths. With the increase of the length, the polar radii $r_0$ and $r_1$ at the start-point and the end-point of Archimedes' helix gradually decrease and increase, respectively, while the polar angle $α_1$ of Archimedes' helix increases significantly. In addition, the failure indices of the Bi-DCBs with different lengths do not reach 1 when they are in the folding stable state, indicating that the Bi-DCBs can achieve the bistable function without failure. With the increase of the length, the Tsai-Hill failure index and the maximum stress failure index almost do not change.

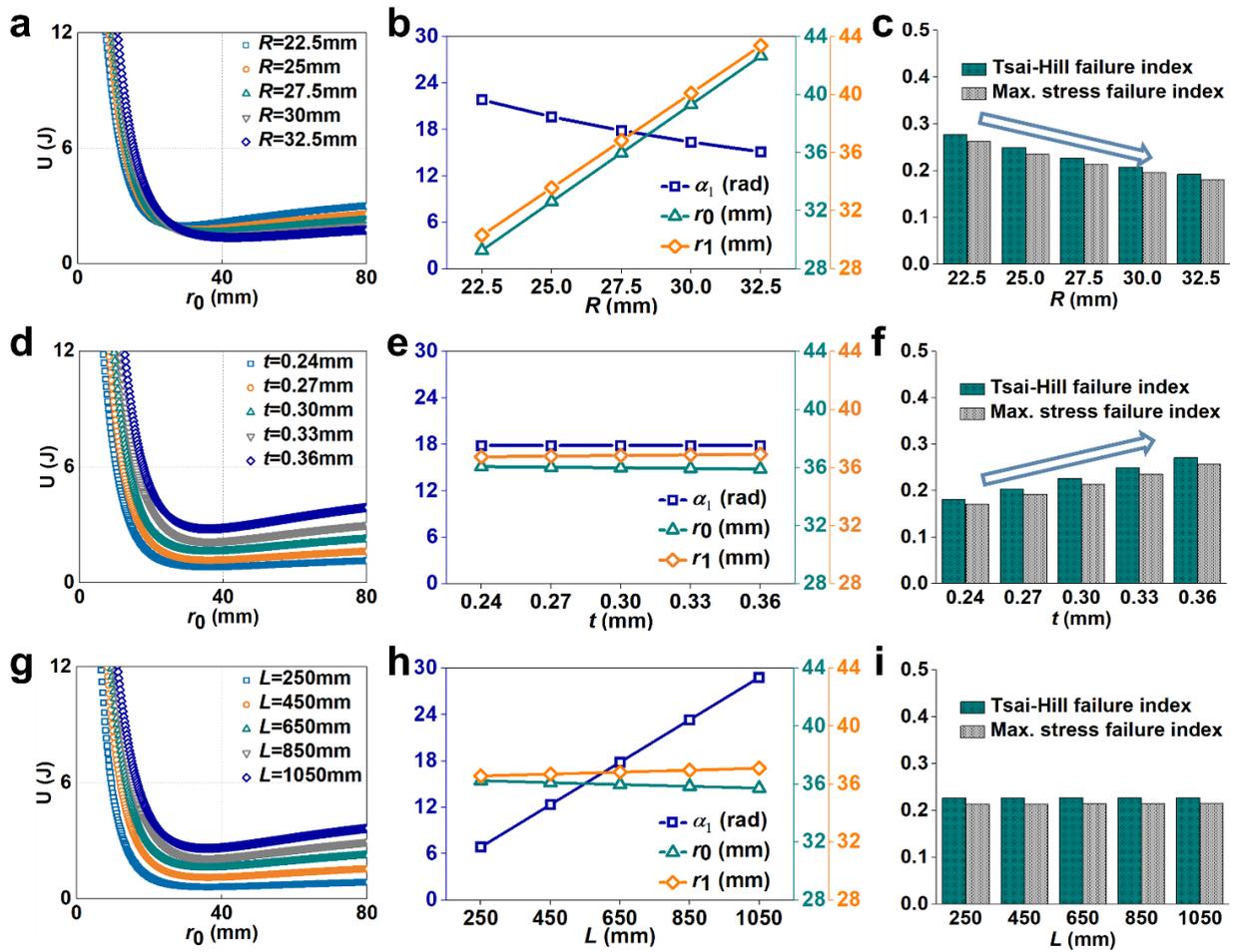

Fig. 8 Effect of geometric parameters on the folding stable state of the Bi-DCB: (a) Total strain energy versus polar radius at the start-point curves with different radii of the cross-section (b) Geometric configuration with different radii of the cross-section (c) Failure indices with different radii of the cross-section (d) Total strain energy versus polar radius at the start-point curves with different thicknesses (e) Geometric configuration with different thicknesses (f) Failure indices with different thicknesses (g)Total strain energy versus polar radius at the start-point curves with different lengths (h) Geometric configuration with different lengths (i) Failure indices with different lengths.

## 6. Conclusions

This paper established an analytical model for predicting the folding stable state of the Bi-DCB, revealing functional mechanisms of the Bi-DCB. Four important results emerging from the research are as follows:

(1) Based on Archimedes' helix equation, the geometric behaviour of the Bi-DCB in the folding stable state was established. Using the energy principle, an analytical model for predicting the folding stable state of the Bi-DCB was derived.

(2) The maximum failure indices of the Bi-DCB in the folding stable state were analyzed by utilizing the Tsai-Hill criterion and the maximum stress criterion.

(3) To validate the analytical model proposed in this paper, the prediction results of the analytical model were compared with experimental results and the results of the two FEMs. The results obtained from the four methods are in good agreement with each other.

(4) With the aid of the analytical model presented in this paper, the influence of geometric parameters (i.e., the radius of the cross-section, thickness and length) on the folding stable state of the Bi-DCB was analyzed.

**Credit authorship contribution statement**

Tian-Wei Liu: Investigation, Conceptualization, Data Curation, Methodology, Software, Formal analysis, Validation, Writing-original draft. Jiang-Bo Bai: Conceptualization, Supervision, Resources, Funding acquisition, Project administration, Methodology, Writing-review & editing. Nicholas Fantuzzi: Validation, Supervision, Writing-review & editing.

**Acknowledgements**

This project was supported by the National Natural Science Foundation of China (Grant No. 52275231 and Grant No. 51875026) and the National Defense Basic Scientific Research Program of China (Grant No. JCKY2019205C002). The first author acknowledges the China Scholarship Council (Grant No. 202106020152).